# Vortex dynamics and pinning properties analysis of $MgB_2$ bulk samples by ac susceptibility measurements

C. Senatore[1], M. Polichetti[1], D. Zola[1], T. Di Matteo[1], G. Giunchi[2], S. Pace[1]

[1]Dipartimento di Fisica "E.R.Caianiello" and INFM, Università di Salerno, Baronissi, Italy
[2] EDISON S.p.A., Milano, Italy

## Abstract

The flux lines dynamics have been investigated on $MgB_2$ bulk superconductors obtained by reactive liquid infiltration by measuring the ac magnetic susceptibility. The fundamental and third harmonics have been studied as a function of temperature, dc magnetic field, ac field amplitude and frequency. In order to determine the dynamical regimes governing the vortex motion, the experimental results have been compared with susceptibility curves obtained by numerical calculations of the non-linear diffusion equation for the magnetic field. The frequency behaviour of the third harmonic response, that cannot be explained by frequency dependent critical state models, has been related to the current dependence of the flux creep activation energy $U(J)$ in the diffusion coefficient. In this way we have shown that the measured curves are correctly interpreted within the framework of a vortex glass description.

*Contacting author*: Carmine Senatore (e-mail: carmelo@sa.infn.it)

## 1. Introduction

The discovery of superconductivity in $MgB_2$ with a transition temperature near to 39 K has led a considerable interest in this binary compound, since there are many proposal for its potential uses exploiting its intrinsic properties. However, the low value of the irreversibility field and the rapid decrease of the critical current density $J_c$ with the magnetic field [1] actually represent a limitation for the high field power applications of $MgB_2$. Many attempts to improve the transport properties by way of proton irradiation [2], alloying and nanoparticles additions [1,3,4] have been done but the results, promising though, are still not conclusive. However, to overcome the present limits of this compound it is necessary to fully understand its pinning properties.

One of the most popular means of investigating the vortex dynamics as well as the pinning properties in type-II superconductors is the measurements of the response of the vortex lattice to the ac magnetic fields [5]. The ac susceptibility represents a powerful tool to study the different dynamical regimes of the flux lines: this technique allows to induce changes in the vortex dynamic by changing the value of the external parameters, such as the ac field frequency and amplitude, the dc field intensity, the temperature, etc. Moreover, in addition to the first harmonic response, the higher order harmonics of ac susceptibility can provide much insights into the loss mechanisms and the flux motion. In fact the response of type-II superconductors to the ac magnetic field can be either linear or non-linear. The linear response can be divided into two different frequency regimes. At high frequency, the response is dominated by a viscous motion of the vortex liquid (flux flow regime) [6]. In the region of extremely low frequency, thermally activated vortex jumps between most favourable metastable states of the vortex lattice come into play (TAFF regime) [7] and contribute to the ac response, especially near $T_c$. In both these cases the linear response results from an Ohmic resistive state $E = r\,J$ in the sample, where $r(B,T)$ is independent of $J$. Therefore, in presence of a dc field larger than the ac one, the electrodynamics of a superconductor in these regimes can be described in terms of a normal metal, governed by skin effect. On the contrary, at intermediate frequencies, when the vortex motion is governed by relevant thermally activated flux creep phenomena, the response of the system is non-linear and the resistivity of the sample is expressed by the $J$-dependent thermally activated form:

$$r(J) = r_0 \, exp\left[ -\frac{U(J)}{k_B T} \right] \tag{1}$$

with $U(J)$ the effective activation energy. In this conditions the response to ac fields cannot be described by analytical approach.

The aim of this work is to relate the frequency behaviour of the third harmonic response $c_3$ of ac susceptibility to the current density dependence of the activation energy $U(J)$ in order to individuate the dynamical regimes governing the vortex motion. In order to do this, we have compared the experimental $c_3(T)$ curves measured for $MgB_2$ bulk superconductors with the simulated behaviour obtained by the numerical integration of the magnetic field diffusion equation for different functional forms of $U(J)$.

This paper is organized as follows. In Sec.2 we report the numerical method applied to resolve the flux diffusion equation. The fundamental and third harmonic measurements performed on $MgB_2$ bulk samples have been discussed and compared with the calculated curves in Sec.3. Finally Sec.4 is dedicated to a summary of this work.

## 2. Numerical method

Since the harmonic susceptibilities are determined by the magnetic flux entering and leaving the sample, we need to study the non-linear diffusion-like equation that governs the spatial-temporal evolution of the local magnetic field in a superconductor. In the case of an infinite slab of thickness $d$ in a parallel field geometry this equation can be written as [8]:

$$\frac{\partial B}{\partial t} = \frac{\partial}{\partial x}\left[ \frac{r(B,T,J)}{m_0} \frac{\partial B}{\partial x} \right] \tag{2}$$

where $r(B, T, J)$ is the residual resistivity due to the flux motion. The proper boundary conditions are $B(\pm d/2, t) = B_{ext}(t) = B_{dc} + B_{ac} \, sin(2p \, f \, t)$.

Equation (2) has been numerically solved by means of the Fortran NAG [9] routines. The algorithm computes the time evolution of the local field profile by integrating a discrete version of Eq.(2) using the Gear's method for a fixed number of spatial meshes. In order to obtain the harmonic susceptibilities $c_n = c'_n + ic''_n$, first we have calculated the magnetization $M$ for the applied time-dependent field, i.e. the magnetization loop, and then its Fourier transforms.

The diffusion coefficient $r(B, T, J)/m_0$ is related to the thermally activated flux motion, leading to electric fields of the form:

$$E = r_n \frac{B}{B_{c2}(T)} J_c(B,T) exp\left[ -\frac{U_P(B,T,J)}{k_B T} \right]. \tag{3}$$

We assume that the temperature, magnetic field and current density dependence in the effective energy barrier $U_P(B, T, J)$ may be separated [10], using for the current dependence the general form [11]

$$U(J) = \frac{U_0}{m} \left[ \left( \frac{J_c}{J} \right)^m - 1 \right]. \tag{4}$$

Using this general expression, different dynamic behaviours, related to the current dependence of the activation energy, can be analysed. In particular, both the vortex glass (VG) scenario (0 £ $m$ £ 1) [12] and the Kim-Anderson (KA) model ($m$ = -1) [13] can be recovered from it. Although experimentally $m$ has been found in literature to depend on both temperature and field [14], in the VG model it is regarded as a universal exponent with a single value, and moreover the choice $m$ =*const*. correctly describes the measured data for a wide range of current densities [15,16]. For this reason, we performed our calculation by considering different constant values of $m$ = 0.1, 0.2, 0.5, 0.8 and -1, thus both in the VG and KA picture. In particular, in the framework of the VG approach for brevity we report here only the curves calculated for the value of the exponent $m$ =0.5.

In order to account for the temperature dependence of the ac susceptibilities, we have to specify the temperature dependences of the critical current density $J_c(T)$ and the activation energy $U_0(T)$. On this purpose, we have adopted the expressions predicted by the collective pinning (CP) model [7,17]:

$$U_P(T) = U_0 \left[ 1 - (T/T_c)^4 \right]$$
$$J_c(T) = J_0 \left[ 1 - (T/T_c)^2 \right]^{5/2} \left[ 1 + (T/T_c)^2 \right]^{-1/2}. \tag{5}$$

Since the simulated curves reported in this paper have been calculated for $B_{dc} >> B_{ac}$ ($B_{ac}$=4 G, $B_{dc}$=200 G), we have neglected the dependence of both $J_c$ and $U_0$ on the ac magnetic field.

The upper critical field has been written as $B_{c2}(T)= B_{c2}(0)\ [1-(T/T_c)^2]/\ [1+(T/T_c)^2]$ [18]. The material parameters used for all calculations presented in this work pertain to a $MgB_2$ slab of thickness $d$=0.5 mm, $T_c$=38.8 K, $B_{c2}(0)$=15 T, $U_0(B_{dc}$=200 G$)/k_B$=8000 K, $J_0(B_{dc}$=200 G$)=5\times10^8$ A/m$^2$, $r_n=2\times10^{-5}$ Ωm.

## 3. Experimental results and discussion

We performed our ac susceptibility studies on $MgB_2$ bulk samples prepared by reactive liquid infiltration [19]. The temperature dependencies of both the first and the third harmonics have been measured changing the temperature with a rate $\Delta T/\Delta t$ = 0.4 K/min. from 4.2 K to 45 K, using different ac field amplitudes (1, 4, 8, 16 G) in the frequency range 27÷3507 Hz, with a dc field $B_{dc}$ ranging from 0 to 200 G. The measurements reported in this paper have been performed on a sample whose dimensions are $20\times2.9\times0.4$ mm$^3$. Both the ac and dc field are applied parallel to the length of the sample. It exhibits a quite sharp transition with a $T_c$=38.8 K, evaluated by the real part of $c_1$ measured for $B_{ac}$=1 G and $f$ =3507 Hz; the transition width, estimated by the 10%-90% criterion, is about 1 K.

The measured $c''_1(T)$ curves reported in Fig.1 show the increase of both the temperature and the amplitude of the peak with the frequency. Also in the $|c_3|(T)$ measurements (Fig.2) a pronounced frequency dependence appears. In fact, by increasing the frequency not only the position of the peak moves toward higher temperatures, but also its height decreases.

Since the experimental data show clear frequency dependences, a simple critical state description [20] is not suitable. Indeed, in the framework of the Bean model, the harmonic susceptibilities can be reduced to universal curves that describe the ac response of a hard superconductor as a function of a single parameter δ, which is the ratio between the full penetration field and the amplitude of the applied alternating field [21]. Therefore, although the introduction of a frequency dependent critical current density might explain the temperature shift of the peak of the harmonic susceptibilities, due to the increase of the current induced by an higher ac field frequency, the same model is not suitable to explain the change in the peak heights which, in this picture, have a defined and constant value.

On the contrary, considering the presence of vortex dynamical regimes and by numerically integrating the magnetic diffusion equation, it has been shown that the increase of the amplitude of $c''_1(T_P)$ with the frequency is an indication of the reduction of the non-linear behaviour of the I-V characteristic driven by the increased electric field [22]. However, to investigate more in detail the flux dynamic we have focused our analysis on the third harmonic components, due to their higher sensitivity to the changes in the dynamical regimes. In particular, from the measurements at increasing frequencies it is clear that the ratio between the heights of the $c'_3(T)$ peaks (the positive one and the negative one), in absolute value, changes. Moreover, our numerical calculations show that, in the KA scenario both the heights of the positive and negative peaks grow up (Fig.3a), whereas in the VG approach the height of both the peaks decreases (Fig.3b). In this way, the frequency dependence of the $c'_3(T)$ curves can provide a valid criterion to distinguish the current dependence form of the activation energy $U(J)$ in the sample. The comparison between the experiments and the simulations shows that the behaviour encountered in the VG approach is the one which better agrees with the frequency dependence of the measured $c'_3(T)$ curves, as it is visible in Fig.4 for the particular values $B_{ac}$=4 G and $B_{dc}$=200 G. As a support of this agreement, it can be noted that the features reported in Fig.2, namely the decrease of $|c_3|(T)$ for increasing frequencies, are also predicted by the numerical simulations in the VG approach [23], as well as it is the reduction of the non-linearity, evidenced from the $c''_1(T)$ curves in Fig.1. Therefore, our results make possible to state that the experimental ac magnetic response of our $MgB_2$ samples is governed by a vortex glass dynamic regime, likely due to the presence of quenched random point-size disorder [19] which promotes the highly disordered glass phase in the vortex matter, as already reported in literature for this kind of material [24].

## 4. Conclusions

In this paper, we have analysed the frequency dependencies of the ac susceptibilities $c_1$ and $c_3$ on $MgB_2$ bulk samples, prepared by reactive liquid infiltration. In particular, by comparison with the numerical calculations of the flux diffusion equation, we have shown that our experimental results represent a strong indication about the presence of a vortex glass dynamic governing the ac magnetic response of the samples.

## Acknowledgements

We want to thank Mr. L. Falco and Mr. A. Ferrentino for their technical support.

**Figure captions**

**Figure 1**: Temperature dependence of $c''_1$ measured at $B_{ac}$=4 G and $f$ =27, 107, 1007, 1607, 2507, 3507 Hz.

**Figure 2**: Temperature dependence of $|c_3|$ measured at $B_{ac}$=4 G and $f$ =27, 107, 1007, 1607, 2507, 3507 Hz.

**Figure 3**: Temperature dependence of $c'_3$ curves calculated for $B_{ac}$=4 G, $B_{dc}$=200 G and $f$ =1007, 1607, 2507, 3507 Hz by using (a) the KA $U(J)$ dependence, (b) a VG-type $U(J)$ dependence in the diffusion coefficient. Insets: Frequency behaviour of the negative peak (open squares) and the positive peak (solid circles) for the calculated $c'_3$ curves.

**Figure 4**: Temperature dependence of $c'_3$ curves measured at $B_{ac}$=4 G, $B_{dc}$=200 G and $f$ =1007, 1607, 2507, 3507 Hz. Insets: Frequency behaviour of the negative peak (open squares) and the positive peak (solid circles) for the measured $c'_3$ curves.

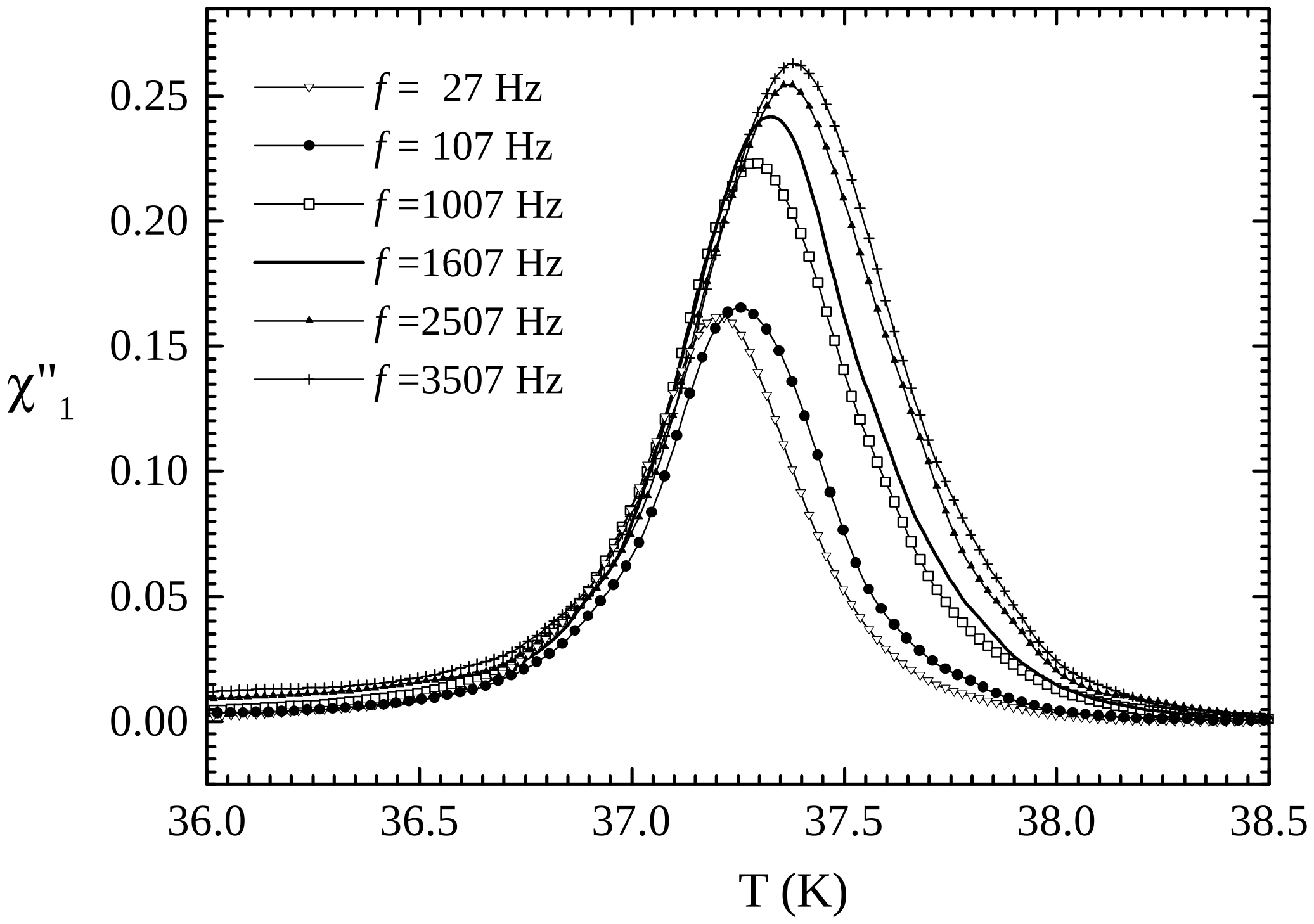

Figure 1

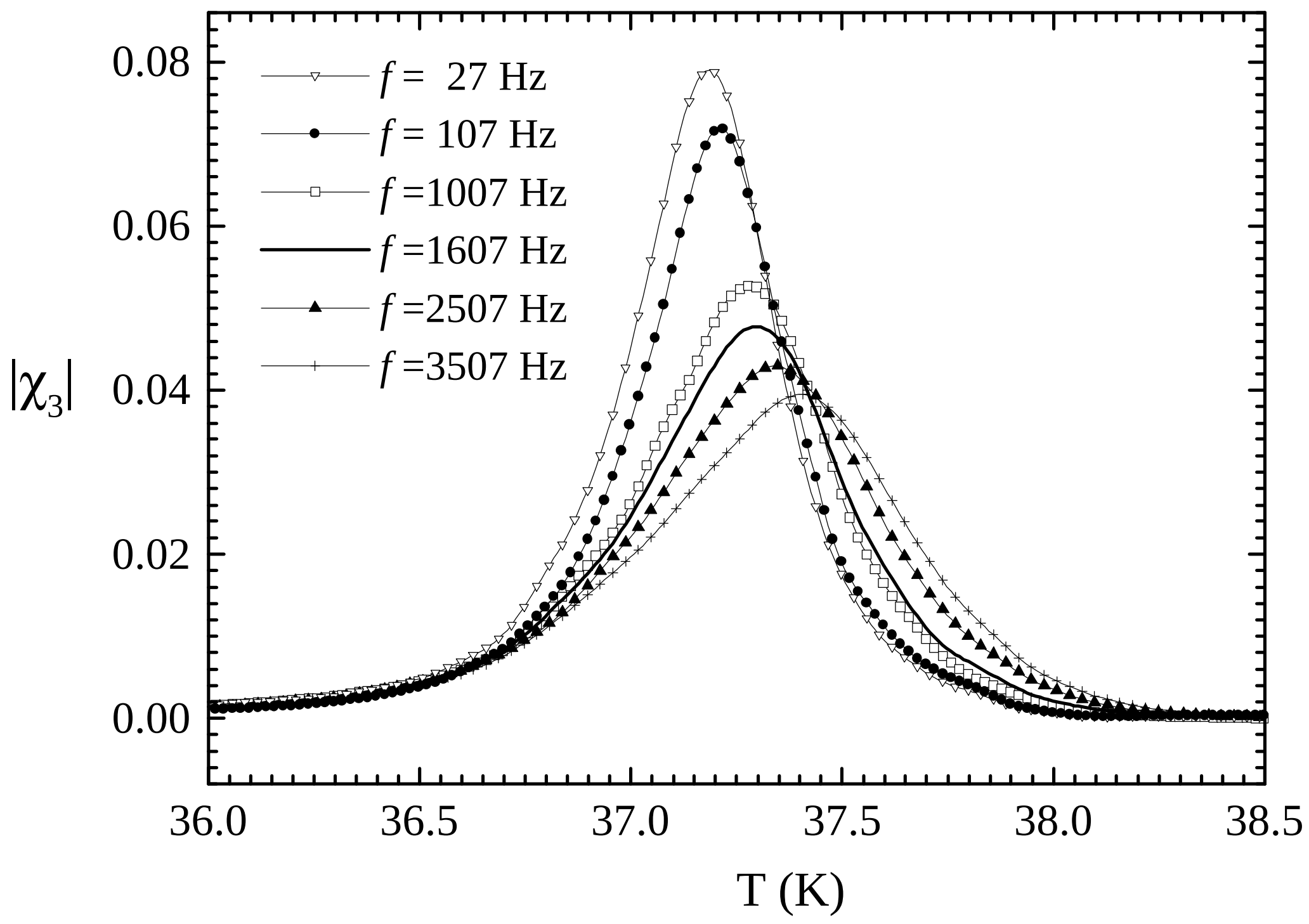

0.08
0.06
0.04
0.02
0.00
|χ3|
f = 27 Hz
f = 107 Hz
f =1007 Hz
f =1607 Hz
f =2507 Hz
f =3507 Hz
36.0
36.5
37.0
37.5
38.0
38.5
T (K)

Figure 2

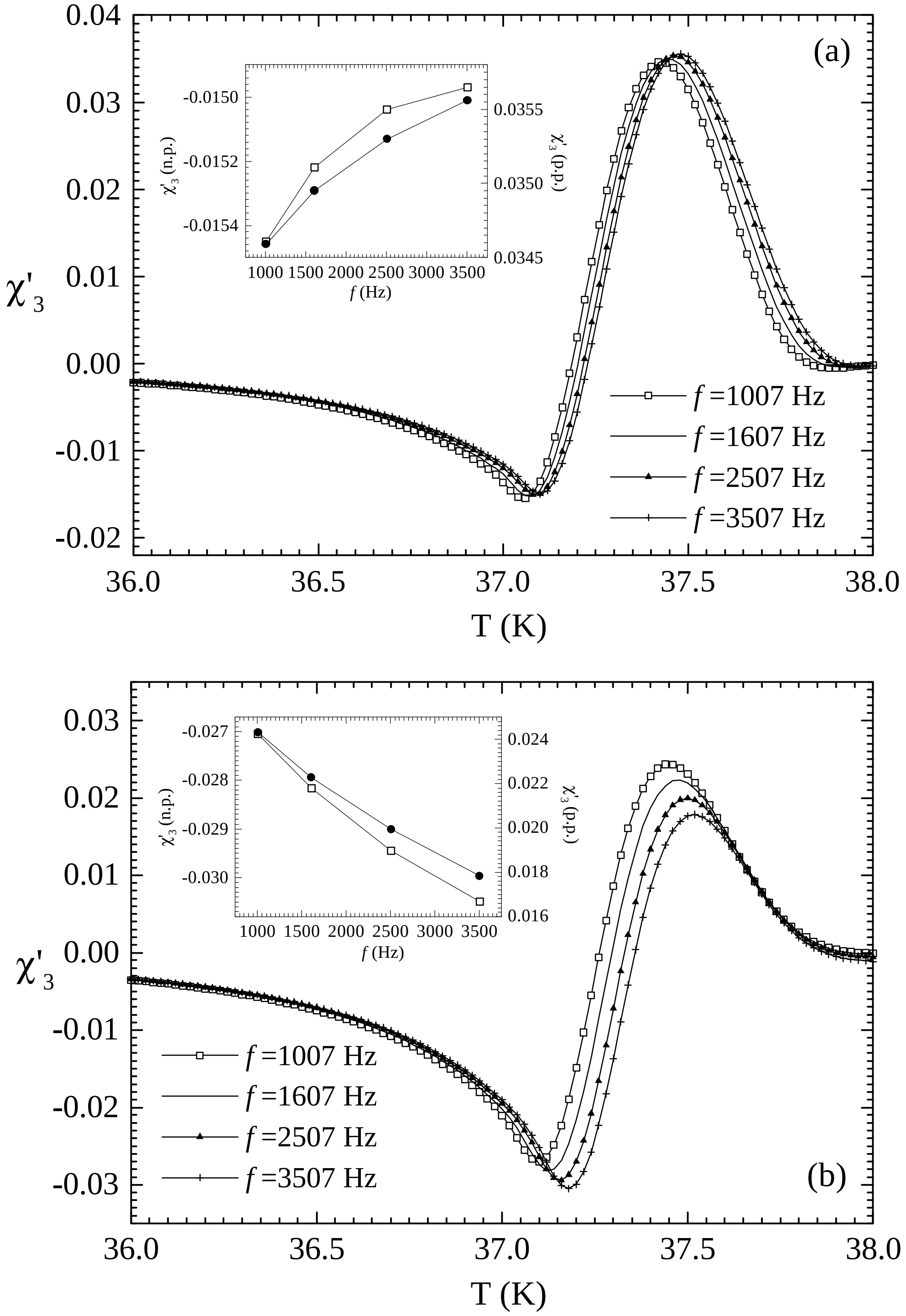

Figure 3

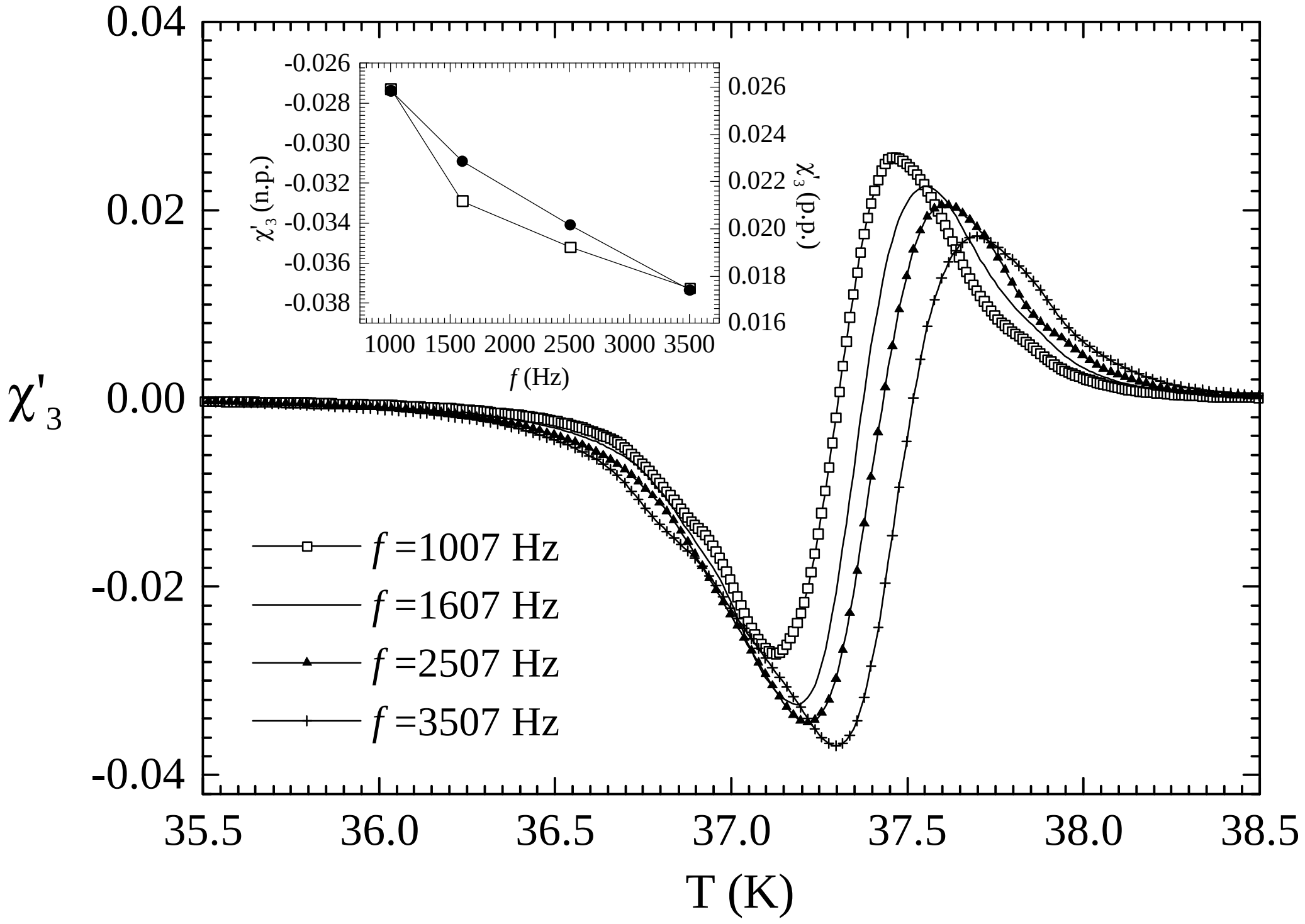

Figure 4